\documentclass[sigconf]{acmart}
\renewcommand\footnotetextcopyrightpermission[1]{} 
\usepackage{caption}

\usepackage{subcaption}
\usepackage{siunitx}
\AtBeginDocument{%
  \providecommand\BibTeX{{%
    \normalfont B\kern-0.5em{\scshape i\kern-0.25em b}\kern-0.8em\TeX}}}

\acmYear{2021}\copyrightyear{2021}
\setcopyright{rightsretained}
\acmConference[WiSec '21]{14th ACM Conference on Security and Privacy in Wireless and Mobile Networks}{June 28--July 2, 2021}{Abu Dhabi, United Arab Emirates}
\acmBooktitle{14th ACM Conference on Security and Privacy in Wireless and Mobile Networks (WiSec '21), June 28--July 2, 2021, Abu Dhabi, United Arab Emirates}
\acmPrice{15.00}
\acmDOI{10.1145/3448300.3468254}
\acmISBN{978-1-4503-8349-3/21/06}

\begin{document}

\title{POSTER: Detecting GNSS misbehaviour with high-precision clocks}

\author{Marco Spanghero}
\orcid{0000-0001-8919-0098}
\affiliation{%
  \institution{Networked Systems Security group}
  \institution{KTH Royal Institute of Technology}
  \city{Stockholm}
  \country{Sweden}
  }
\email{marcosp@kth.se}

\author{Panos Papadimitratos}
\orcid{0000-0002-3267-5374}
\affiliation{%
\institution{Networked Systems Security group}
  \institution{KTH Royal Institute of Technology}
  \city{Stockholm}
  \country{Sweden}
}
\email{papadim@kth.se}


\begin{abstract}

To mitigate spoofing attacks targeting global navigation satellite systems (GNSS) receivers, one promising method is to rely on alternative time sources, such as network-based synchronization, in order to detect clock offset discrepancies caused by GNSS attacks. However, in case of no network connectivity, such validation references would not be available. A viable option is to rely on a local time reference; in particular, precision hardware clock ensembles of chip-scale thermally stable oscillators with extended holdover capabilities. We present a preliminary design and results towards a custom device capable of providing a stable reference, with smaller footprint and cost compared to traditional precision clocks. The system is fully compatible with existing receiver architecture, making this solution feasible for most industrial scenarios. Further integration with network-based synchronization can provide a complete time assurance system, with high short- and long-term stability.
\end{abstract}

\maketitle
\pagestyle{plain}

\section{Introduction}
Ubiquitous GNSS receivers provide precise location and time to a wide gamut of applications, beyond navigation, for mobile communication systems. The inherent vulnerability of GNSS signals allows attackers to produce adversarial signals that purposely alter the GNSS receiver position, velocity, and timing (PVT) solution. Broad availability of affordable Software Defined Radios (SDRs) and open-source tools for GNSS spoofing/meaconing pose a significant threat to GNSS receivers deployed in critical applications \cite{humphreys2008assessing, zeng2018all, Shepard2012EvaluationAttacks}. Authentication methods, such as the Galileo Navigation Message Authentication (NMA), do not fully addresses the problem \cite{walker2015galileo}, as they do not preclude attacks that relay/replay legitimate signals \cite{PapadimitratosJa:C:2008, zhang2019safe, zhang2019on}. Moreover, NMA is not backwards compatible with existing receiver hardware and requires changes to the structure of the signal in space. Several countermeasures were proposed, leveraging signal characteristics, e.g. \cite{Borio2021GNSSAssessment}, the receiver's attitude \cite{Broumandan2018SpoofingNavigation} or validation of the PVT solution through alternative position and time sources \cite{PapadimitratosJ:P:2012, Oligeri2019DriveExperiments, spangheroGNSS20, kzmsppPLANS2020}. 

Attackers tampering with the PVT solution produce noticeable effects on the GNSS receiver clock, independently of their specific target. Countermeasures based on the observation of the receiver clock state could be agnostic to the exact attack form deployed or the attacker's objective. 
Considering the attack detection, intuitively, it is reasonable to compare the progression of time in the GNSS receiver against a trusted reference (e.g., remote time servers) to monitor  for changes that would hint to an adversarial action. In fact, modern receivers are rarely used as stand-alone devices, and they are often integrated into complex devices with different types of network connectivity, allowing fusion with other sensor- and Internet-provided data. 

On the other hand, the receiver might be prevented from accessing network based synchronization services for extended periods of times (e.g., scarce network coverage or even adversarial denial of service). The fall-back approach for the GNSS-enabled system is to rely on on-board crystal oscillators. Performing a so-called 'time test' is not a new idea  \cite{Arafin2017, marnach2013detecting, Papadimitratos2008GNSS-basedCountermeasures}, but integration of cheap clock references with the GNSS receiver is challenging due to environment-dependant effects, such as temperature variations, or the precision of the oscillator itself. To overcome some of these limitations, one could rely on improved reference clocks (e.g., oven- or double oven-compensated oscillators or rubidium references). But this comes at an increase in cost, footprint and power consumption, making such an not feasible for many mobile platforms.

Ultimately, an option is to use ensembles of multiple, local, chip-scale high precision hardware reference oscillators. This approach compensates for errors caused by manufacturing inaccuracies, thermal deviations or frequency instability that affect single references. In \cite{Greenhall2011ReducedEnsembles}, an optimal method based on Kalman filters was presented, to produce clock ensembles, improving the long-term stability over a single reference up to one order of magnitude. By adopting stability improvement mechanisms, our aim is to provide a stable clock reference the receiver can use to evaluate the state of the GNSS time information and by extension detect GNSS attacks.

This work investigates the feasibility of a solution based on precision chip-scale clocks to provide a detection (and recovery) system from GNSS attacks (spoofing, replaying/relaying). Initial results based on our system, using a single-precision chip-scale, oven-compensated oscillators (OCXOs) are presented. The device is tested in a realistic scenario, using the Texas Spoofing Battery (TEXBAT) trace files \cite{humphreys2012texas} to spoof a commercial GNSS receiver. Additionally, we present a preliminary design with multiple oscillators, to achieve a low-power, high-stability clock ensemble that can be used to provide cost-effective (compared to existing time assurance systems based on chip-scale atomic clocks), high-performance reference for GNSS attack/fault detection.

\begin{figure*}
  \includegraphics[width=0.6\textwidth]{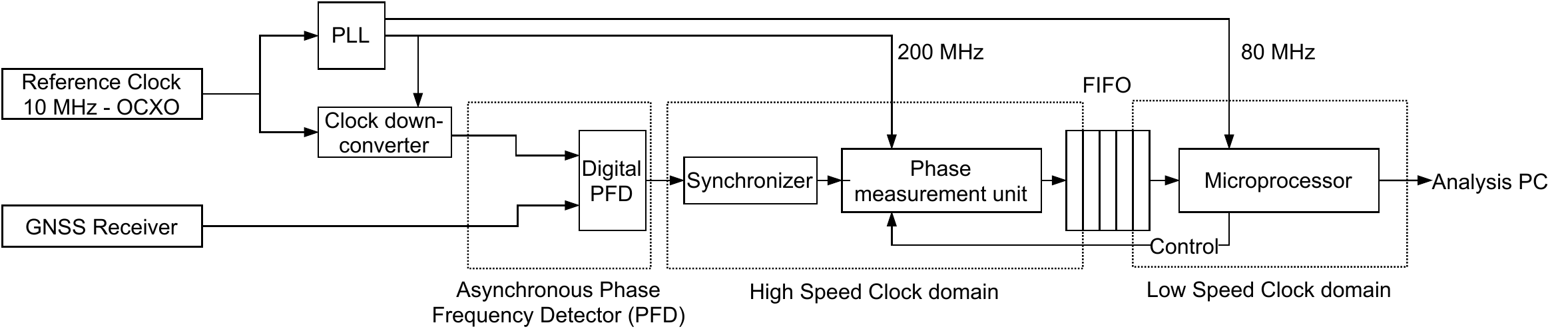}
  \caption{Block diagram of the system: the phase offset is used to detect potential GNSS attacks.}
  \Description{Block diagram of the system. The phase offset between GNSS PPS and clock reference is used to detect potential GNSS attacks.}
  \label{fig:block-diagram}
\end{figure*}

\section{Design}
GNSS receivers can benefit from high-precision stable clocks to improve their performance. Ultra-stable clocks allow GNSS receivers to operate with a reduced number of satellites, providing a PVT solution even if less than four satellites are in view; although with some limitations for extremely long integration times \cite{Gaggero2008Ultra-stableIntegration}. On the other hand, such stable clocks are expensive, bulky, and often not suited for embedded computing devices or many relatively small-footprint mobile devices. Low-cost oscillators can be used, but they have poor stability and require continuous tuning. For this reason, they are not stable enough for attack detection. In contrast, chip-scale OCXOs provide adequate performance as clock references, in a compact and power-efficient form factor. 

This is why we develop a custom platform based on commercially available components, designed to work in tight integration with the GNSS receiver. The clock states (phase offset, frequency offset and frequency drift) of the GNSS receiver are continuously tested against the custom reference, to detect misbehavior effects on the GNSS receiver clock. The design estimates the relative phase between the 1-Pulse-per-Second (PPS) signal (which is synchronous to the top of the second of the GNSS PVT solution) against a 1-PPS clock obtained from a free-running hardware clock. Any deviation of the GPS time induced by the attacker causes a phase change to the victim receiver PPS. If such variation is beyond the normal drift rate of the reference clock, it can be an indication of an attack. 

The device is implemented with mixed clock domain blocks in a Field Programmable Gate Array (FPGA). The FPGA implementation allows for high-resolution measurements of the phase deviation, with predictable latency. Figure \ref{fig:block-diagram} shows the implementation of the detection and control logic. The reference high-precision, chip-scale OCXO is down-converted to produce the 1-PPS reference signal, and up-converted with a Phase-Locked Loop (PLL) clock synthesizer to provide the fast measurement clock. The reference OCXO has a frequency of \SI{10}{\mega\hertz}, the measurement clock derived from the PLL is \SI{200}{\mega\hertz}, allowing a resolution of \SI{5}{\nano\second}.

To precisely track the phase offset, a Kalman filter is designed as described in \cite{Greenhall2001KalmanAlgorithm}. Due to its low computational complexity, the filter can be computed on embedded hardware. At this stage, for development, the computation is offloaded to the acquisition laptop.

\section{Experimental setup}
The proposed design is implemented on an Intel Altera MAX10 FPGA. The phase detection and measurement system is implemented in hardware and validated up to a frequency of 200Mhz. The receiver tested is a u-Blox C099-F9P high-performance dual-band receiver, configured to operate on a single frequency of the GPS constellation only, to comply to the TEXBAT scenarios. The PPS frequency is \SI{1}{\hertz}, commonly used in industrial applications. 

\begin{figure}
  \includegraphics[width=0.70\linewidth]{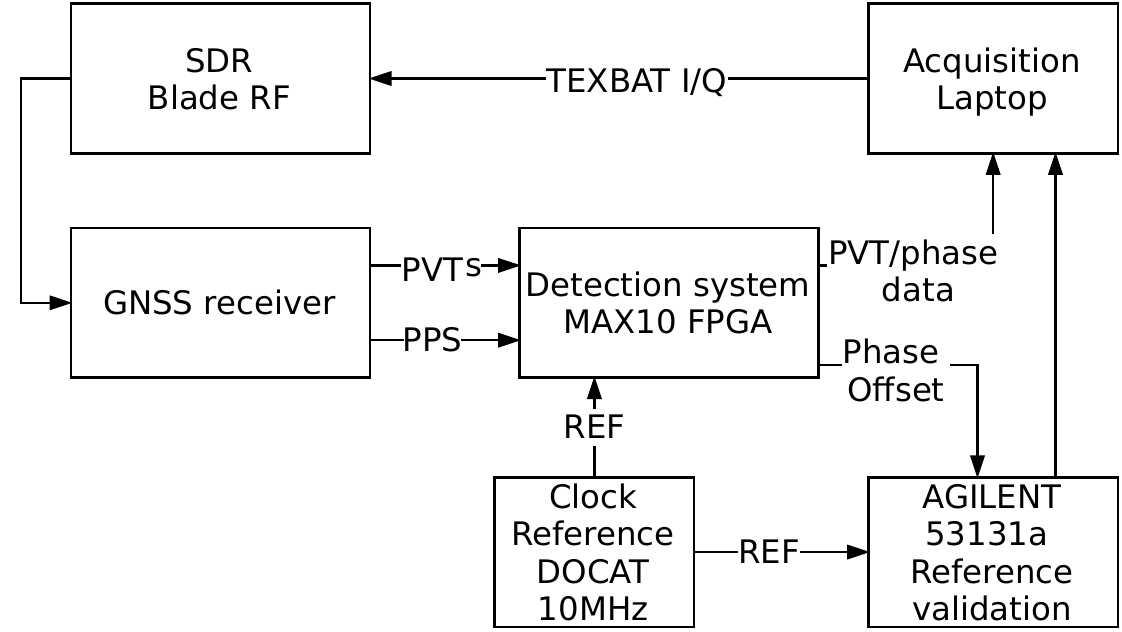}
  \caption{Experimental setup used to test the design. The Agilent counter measurements are used to validate the system stability and accuracy.}
  \label{fig:experimental-diagram}
\end{figure}

The spoofing signals are generated using a Nuand BladeRF SDR, that transmits raw I/Q samples from the TEXBAT scenario under test. 
Figure \ref{fig:experimental-diagram} shows the layout of the experimental testbed. The phase measurements obtained from the MAX10 FPGA are stored on the acquisition computer and validated against the Agilent counter. Both instruments use the same high-precision chip-scale OCXO (Allan deviation $\sigma_{y}(\tau)=\num{8e-11}$) reference to produce comparable measurements. 

\section{Preliminary results}
\begin{figure}
     \centering
     \begin{subfigure}{0.7\linewidth}
         \centering
         \includegraphics[width=\linewidth]{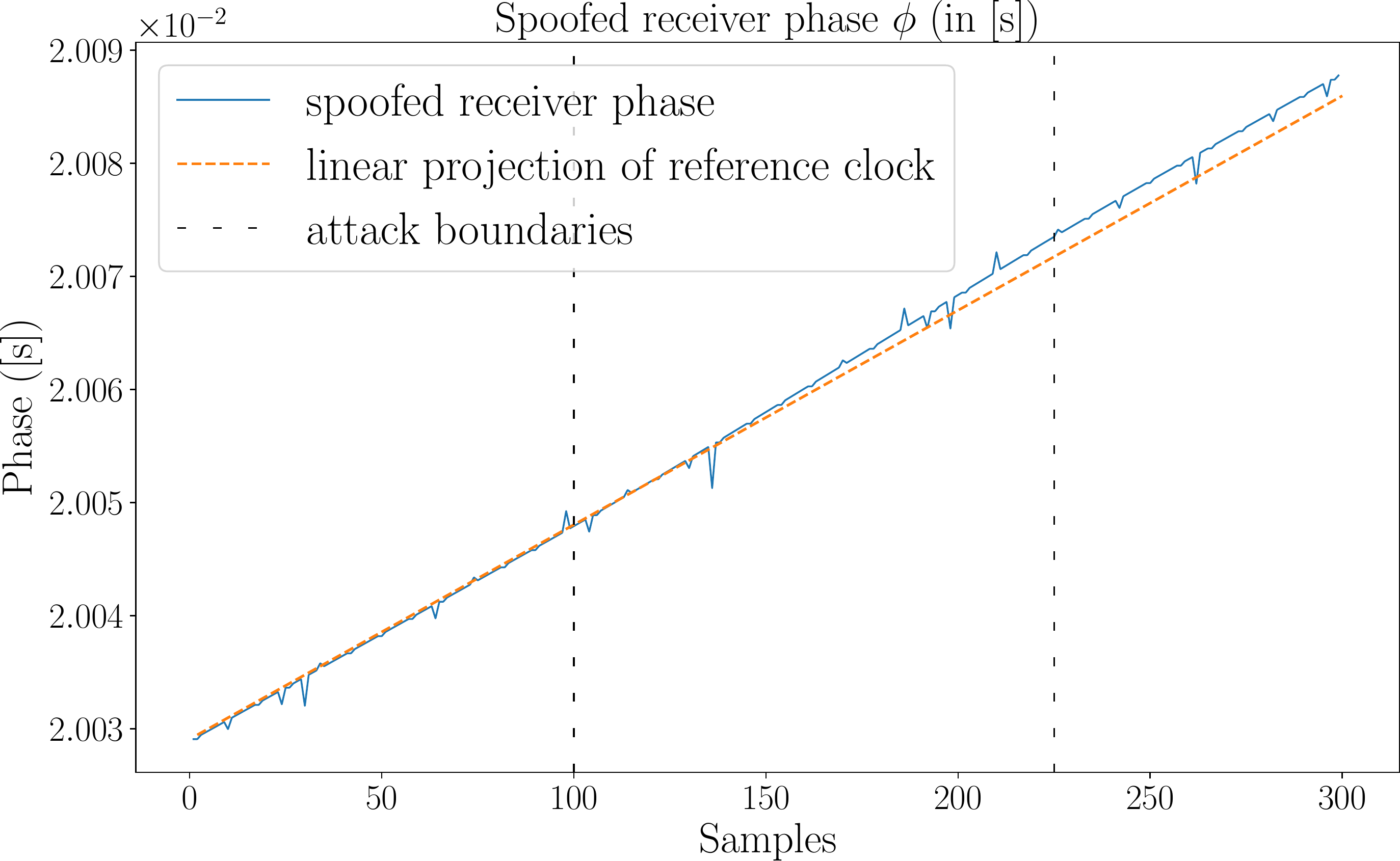}
         \caption{Spoofed receiver PPS phase}
         \label{fig:prelim-res-delta}
     \end{subfigure}
     
     \begin{subfigure}{0.7\linewidth}
         \centering
         \includegraphics[width=\linewidth]{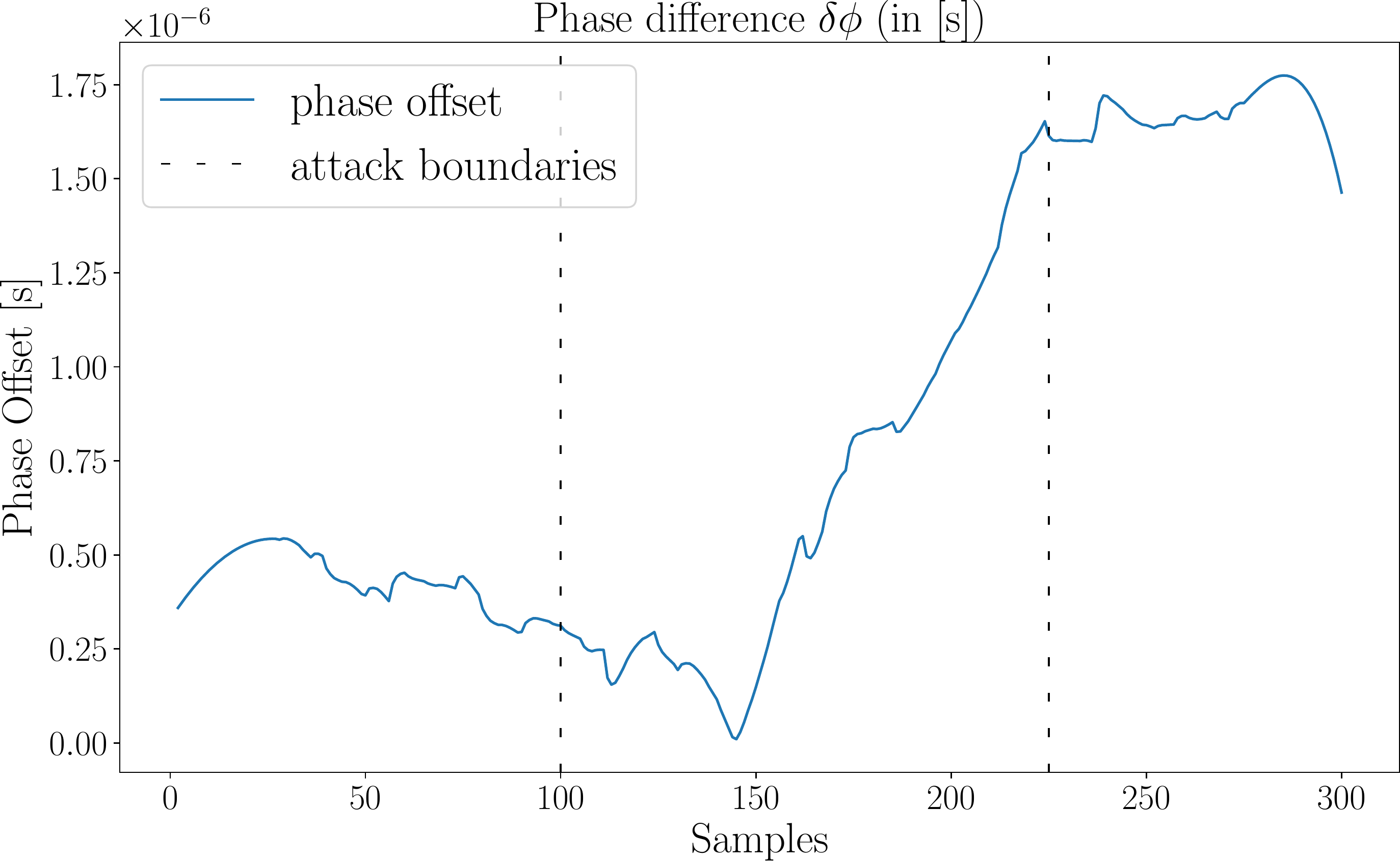}
         \caption{Phase difference of GNSS receiver and reference}
         \label{fig:prelim-res-delta-phase}
     \end{subfigure}
        \caption{Spoofed receiver phase variation against a local high precision reference.}
        \label{fig:prelim-res}
\end{figure}

The proposed system was tested against two time-focused scenarios from TEXBAT (scenarios ds-2 and ds-3). After the receiver obtains a legitimate PVT solution, the attack starts roughly at sample 100; once the phase offset reaches \SI{2}{\micro\second} (Figure \ref{fig:prelim-res-delta}), the target is considered completely captured. Our design detects adversary-induced errors in accordance with \cite{humphreys2012texas}, demonstrating the validity of the presented approach. Figure \ref{fig:prelim-res-delta-phase} shows the abnormal change in the phase offset, as detected by the proposed method, revealing the attacker-induced drift of the GNSS clock. The total deviation allowed to avoid a false positive detection depends on the specific application (e.g., power grid synchrophasors must not drift more than \SI{25.6}{\micro\second} \cite{IEEE/IECMeasurements}).
The stability of our reference clock suggests that the detection threshold could be improved, but initial observations suggest that the phase noise (denoted by the slope of the phase measurements) of the embedded PLL limits the detection factor. The observed phase drift of the PLL output against a rubidium reference is $\approx\SI{90}{\nano\second/\second}$, establishing the lower detection bound for our proposed system.

\section{Conclusions and future work}
The initial design can detect phase changes in the PPS edge against our high-precision reference, with a \SI{5}{\nano\second} resolution, and stability of \SI{90}{\nano\second/\second}. Limitations due to the PLL phase stability were identified and possible solutions are under investigation. 

\begin{figure}
  \includegraphics[width=0.75\linewidth]{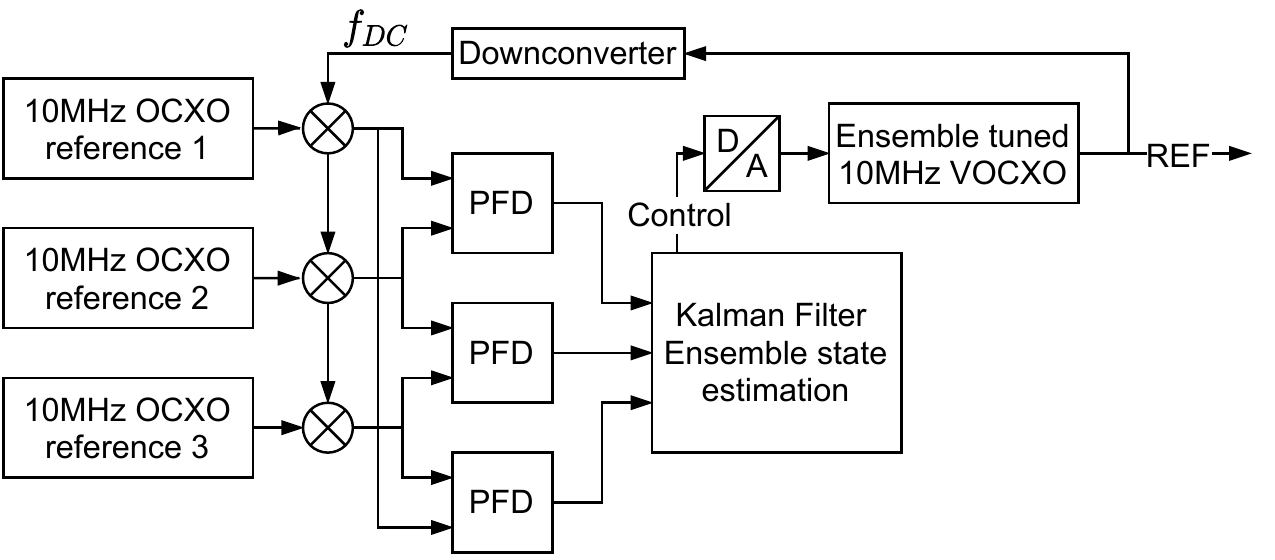}
  \caption{Hardware clock ensemble for enhanced time.}
  \label{fig:prelim-ensemble}
\end{figure}

\textbf{Ongoing work:} A design based on ensembles of multiple precise clock references is under development, with expected stability improvement by an order of magnitude compared to the single-clock design evaluated here. Based on a non-uniform sampling Kalman filter, the ensemble aims at providing a long-term stable reference for detecting GNSS spoofing attacks. Figure \ref{fig:prelim-ensemble} presents a preliminary overview of the hardware ensemble. A group of free-running OCXO clocks is used to fine-tune a voltage-controlled OCXO that provides a system-wide clock for the measurement, estimation and misbehaviour detection. The hardware implementation of such system would provide extended holdover capabilities and result in a computationally efficient and cost-effective solution. This would strengthen GNSS receivers against spoofing attacks. Integration of this system with network-based time provision systems, would provide a complete solution for time based GNSS attack detection.

\begin{acks}
This work was supported by the Swedish Foundation for Strategic Research (SSF) SURPRISE project and the KAW Academy Fellow Trustworthy IoT project.
\end{acks}

\bibliographystyle{ACM-Reference-Format}
\bibliography{stripped-ref-nodoi}


\begin{thebibliography}{21}


\ifx \showCODEN    \undefined \def \showCODEN     #1{\unskip}     \fi
\ifx \showDOI      \undefined \def \showDOI       #1{#1}\fi
\ifx \showISBNx    \undefined \def \showISBNx     #1{\unskip}     \fi
\ifx \showISBNxiii \undefined \def \showISBNxiii  #1{\unskip}     \fi
\ifx \showISSN     \undefined \def \showISSN      #1{\unskip}     \fi
\ifx \showLCCN     \undefined \def \showLCCN      #1{\unskip}     \fi
\ifx \shownote     \undefined \def \shownote      #1{#1}          \fi
\ifx \showarticletitle \undefined \def \showarticletitle #1{#1}   \fi
\ifx \showURL      \undefined \def \showURL       {\relax}        \fi
\providecommand\bibfield[2]{#2}
\providecommand\bibinfo[2]{#2}
\providecommand\natexlab[1]{#1}
\providecommand\showeprint[2][]{arXiv:#2}

\bibitem[\protect\citeauthoryear{??}{IEE}{2018}]%
        {IEEE/IECMeasurements}
 \bibinfo{year}{2018}\natexlab{}.
\newblock \showarticletitle{IEEE/IEC International Standard - Measuring relays
  and protection equipment - Part 118-1: Synchrophasor for power systems -
  Measurements}.
\newblock \bibinfo{journal}{\emph{IEC/IEEE 60255-118-1:2018}}
  (\bibinfo{year}{2018}), \bibinfo{pages}{1--78}.
\newblock


\bibitem[\protect\citeauthoryear{Arafin, Anand, and Qu}{Arafin
  et~al\mbox{.}}{2017}]%
        {Arafin2017}
\bibfield{author}{\bibinfo{person}{Md~T. Arafin}, \bibinfo{person}{D. Anand},
  {and} \bibinfo{person}{G. Qu}.} \bibinfo{year}{2017}\natexlab{}.
\newblock \showarticletitle{{A low-cost GPS spoofing detector design for
  Internet of Things (IoT) applications}}. In
  \bibinfo{booktitle}{\emph{Proceedings of the ACM GLSVLSI}} (Banff, Canada).
\newblock


\bibitem[\protect\citeauthoryear{Borio and Gioia}{Borio and Gioia}{2021}]%
        {Borio2021GNSSAssessment}
\bibfield{author}{\bibinfo{person}{D Borio} {and} \bibinfo{person}{C Gioia}.}
  \bibinfo{year}{2021}\natexlab{}.
\newblock \showarticletitle{GNSS interference mitigation: A measurement and
  position domain assessment}.
\newblock \bibinfo{journal}{\emph{NAVIGATION}} \bibinfo{volume}{68},
  \bibinfo{number}{1} (\bibinfo{year}{2021}), \bibinfo{pages}{93--114}.
\newblock


\bibitem[\protect\citeauthoryear{Broumandan and Lachapelle}{Broumandan and
  Lachapelle}{2018}]%
        {Broumandan2018SpoofingNavigation}
\bibfield{author}{\bibinfo{person}{A. Broumandan} {and} \bibinfo{person}{G.
  Lachapelle}.} \bibinfo{year}{2018}\natexlab{}.
\newblock \showarticletitle{Spoofing Detection Using GNSS/INS/Odometer Coupling
  for Vehicular Navigation}.
\newblock \bibinfo{journal}{\emph{Sensors}} \bibinfo{volume}{18},
  \bibinfo{number}{5} (\bibinfo{year}{2018}).
\newblock


\bibitem[\protect\citeauthoryear{Gaggero and Borio}{Gaggero and Borio}{2008}]%
        {Gaggero2008Ultra-stableIntegration}
\bibfield{author}{\bibinfo{person}{P.~O. Gaggero} {and} \bibinfo{person}{D.
  Borio}.} \bibinfo{year}{2008}\natexlab{}.
\newblock \showarticletitle{{Ultra-stable oscillators: Limits of GNSS coherent
  integration}}. In \bibinfo{booktitle}{\emph{Proceedings of ION GNSS 2008}}
  (Savannah, GA).
\newblock


\bibitem[\protect\citeauthoryear{Greenhall}{Greenhall}{2001}]%
        {Greenhall2001KalmanAlgorithm}
\bibfield{author}{\bibinfo{person}{C.~A. Greenhall}.}
  \bibinfo{year}{2001}\natexlab{}.
\newblock \showarticletitle{{Kalman plus weights: a time scale algorithm}}. In
  \bibinfo{booktitle}{\emph{Proceedings of PTTI}} (Long Beach, CA).
\newblock


\bibitem[\protect\citeauthoryear{Greenhall}{Greenhall}{2011}]%
        {Greenhall2011ReducedEnsembles}
\bibfield{author}{\bibinfo{person}{C.~A. Greenhall}.}
  \bibinfo{year}{2011}\natexlab{}.
\newblock \showarticletitle{Reduced Kalman filters for clock ensembles}. In
  \bibinfo{booktitle}{\emph{Proceedings of Joint Conference of the IEEE
  International Frequency Control and the European Frequency and Time Forum}}
  (San Francisco, CA).
\newblock


\bibitem[\protect\citeauthoryear{Humphreys, Bhatti, Shepard, and
  Wesson}{Humphreys et~al\mbox{.}}{2012}]%
        {humphreys2012texas}
\bibfield{author}{\bibinfo{person}{T.~E. Humphreys}, \bibinfo{person}{J.~A.
  Bhatti}, \bibinfo{person}{D. Shepard}, {and} \bibinfo{person}{K. Wesson}.}
  \bibinfo{year}{2012}\natexlab{}.
\newblock \showarticletitle{{The Texas spoofing test battery: Toward a standard
  for evaluating GPS signal authentication techniques}}. In
  \bibinfo{booktitle}{\emph{Proceedings of ION GNSS}} (Nashville, TN).
\newblock


\bibitem[\protect\citeauthoryear{Humphreys, Ledvina, Psiaki, O'Hanlon, and
  Kintner}{Humphreys et~al\mbox{.}}{2008}]%
        {humphreys2008assessing}
\bibfield{author}{\bibinfo{person}{T.~E. Humphreys}, \bibinfo{person}{B.~M.
  Ledvina}, \bibinfo{person}{M.~L. Psiaki}, \bibinfo{person}{B.~W. O'Hanlon},
  {and} \bibinfo{person}{P.~M. Kintner}.} \bibinfo{year}{2008}\natexlab{}.
\newblock \showarticletitle{{Assessing the spoofing threat: Development of a
  portable GPS civilian spoofer}}. In \bibinfo{booktitle}{\emph{ION GNSS}}
  (Savannah, GA).
\newblock


\bibitem[\protect\citeauthoryear{Marnach, Mauw, Martins, and Harpes}{Marnach
  et~al\mbox{.}}{2013}]%
        {marnach2013detecting}
\bibfield{author}{\bibinfo{person}{D. Marnach}, \bibinfo{person}{S. Mauw},
  \bibinfo{person}{M. Martins}, {and} \bibinfo{person}{C. Harpes}.}
  \bibinfo{year}{2013}\natexlab{}.
\newblock \showarticletitle{{Detecting Meaconing Attacks by Analysing the Clock
  Bias of GNSS Receivers}}.
\newblock \bibinfo{journal}{\emph{Artificial Satellites}} \bibinfo{volume}{48},
  \bibinfo{number}{2} (\bibinfo{date}{May} \bibinfo{year}{2013}),
  \bibinfo{pages}{63--83}.
\newblock
\showISSN{0208-841X}


\bibitem[\protect\citeauthoryear{Oligeri, Sciancalepore, Ibrahim, and
  Di~Pietro}{Oligeri et~al\mbox{.}}{2019}]%
        {Oligeri2019DriveExperiments}
\bibfield{author}{\bibinfo{person}{G. Oligeri}, \bibinfo{person}{S.
  Sciancalepore}, \bibinfo{person}{Omar~A. Ibrahim}, {and} \bibinfo{person}{R.
  Di~Pietro}.} \bibinfo{year}{2019}\natexlab{}.
\newblock \showarticletitle{{Drive me not: GPS spoofing detection via cellular
  network (architectures, models, and experiments)}}. In
  \bibinfo{booktitle}{\emph{Proceedings of ACM WiSec'19}} (Miami, FL).
\newblock


\bibitem[\protect\citeauthoryear{Papadimitratos and Jovanovic}{Papadimitratos
  and Jovanovic}{2008a}]%
        {Papadimitratos2008GNSS-basedCountermeasures}
\bibfield{author}{\bibinfo{person}{P. Papadimitratos} {and} \bibinfo{person}{A.
  Jovanovic}.} \bibinfo{year}{2008}\natexlab{a}.
\newblock \showarticletitle{GNSS-based Positioning: Attacks and
  countermeasures}. In \bibinfo{booktitle}{\emph{MILCOM 2008}} (San Diego, CA).
\newblock


\bibitem[\protect\citeauthoryear{Papadimitratos and Jovanovic}{Papadimitratos
  and Jovanovic}{2008b}]%
        {PapadimitratosJa:C:2008}
\bibfield{author}{\bibinfo{person}{P. Papadimitratos} {and} \bibinfo{person}{A.
  Jovanovic}.} \bibinfo{year}{2008}\natexlab{b}.
\newblock \showarticletitle{{Protection and Fundamental Vulnerability of
  GNSS}}. In \bibinfo{booktitle}{\emph{IEEE IWSSC}} (Toulouse, France).
\newblock


\bibitem[\protect\citeauthoryear{Papadimitratos and Jovanovic}{Papadimitratos
  and Jovanovic}{2012}]%
        {PapadimitratosJ:P:2012}
\bibfield{author}{\bibinfo{person}{P. Papadimitratos} {and} \bibinfo{person}{A.
  Jovanovic}.} \bibinfo{year}{2012}\natexlab{}.
\newblock \bibinfo{title}{{Method to Secure GNSS-based Locations in a Device
  having GNSS Receiver}}.
\newblock
\newblock
\urldef\tempurl%
\url{http://www.google.com/patents/US8159391}
\showURL{%
\tempurl}
\newblock
\shownote{{US} {P}atent 8,159,391.}


\bibitem[\protect\citeauthoryear{Shepard, Humphreys, and Fansler}{Shepard
  et~al\mbox{.}}{2012}]%
        {Shepard2012EvaluationAttacks}
\bibfield{author}{\bibinfo{person}{D.~P. Shepard}, \bibinfo{person}{T.~E.
  Humphreys}, {and} \bibinfo{person}{A.~A. Fansler}.}
  \bibinfo{year}{2012}\natexlab{}.
\newblock \showarticletitle{Evaluation of the vulnerability of phasor
  measurement units to GPS spoofing attacks}.
\newblock \bibinfo{journal}{\emph{International Journal of Critical
  Infrastructure Protection}} \bibinfo{volume}{5}, \bibinfo{number}{3}
  (\bibinfo{year}{2012}), \bibinfo{pages}{146--153}.
\newblock


\bibitem[\protect\citeauthoryear{Spanghero, Zhang, and
  Papadimitratos}{Spanghero et~al\mbox{.}}{2020}]%
        {spangheroGNSS20}
\bibfield{author}{\bibinfo{person}{M. Spanghero}, \bibinfo{person}{K. Zhang},
  {and} \bibinfo{person}{P. Papadimitratos}.} \bibinfo{year}{2020}\natexlab{}.
\newblock \showarticletitle{{Authenticated Time for Detecting GNSS Attacks}}.
  In \bibinfo{booktitle}{\emph{Proceedings of ION GNSS+ 2020}} (Virtual
  Conference).
\newblock


\bibitem[\protect\citeauthoryear{Walker, Rijmen, Fernandez-Hernandez,
  Seco-Granados, Sim{\'{o}}n, Calle, and Pozzobon}{Walker
  et~al\mbox{.}}{2015}]%
        {walker2015galileo}
\bibfield{author}{\bibinfo{person}{P. Walker}, \bibinfo{person}{V. Rijmen},
  \bibinfo{person}{I. Fernandez-Hernandez}, \bibinfo{person}{G. Seco-Granados},
  \bibinfo{person}{J. Sim{\'{o}}n}, \bibinfo{person}{J.~D. Calle}, {and}
  \bibinfo{person}{O. Pozzobon}.} \bibinfo{year}{2015}\natexlab{}.
\newblock \showarticletitle{{Galileo open service authentication: a complete
  service design and provision analysis}}. In
  \bibinfo{booktitle}{\emph{Proceedings of ION GNSS+ 2015}} (Tampa, FL).
  \bibinfo{pages}{3383--3396}.
\newblock


\bibitem[\protect\citeauthoryear{Z., L., S., W., L., D., W., and Y.}{Z.
  et~al\mbox{.}}{2018}]%
        {zeng2018all}
\bibfield{author}{\bibinfo{person}{Kexiong~C. Z.}, \bibinfo{person}{Shinan L.},
  \bibinfo{person}{Yuanchao S.}, \bibinfo{person}{Dong W.},
  \bibinfo{person}{Haoyu L.}, \bibinfo{person}{Yanzhi D.},
  \bibinfo{person}{Gang W.}, {and} \bibinfo{person}{Yaling Y.}}
  \bibinfo{year}{2018}\natexlab{}.
\newblock \showarticletitle{All Your {GPS} Are Belong To Us: Towards Stealthy
  Manipulation of Road Navigation Systems}. In \bibinfo{booktitle}{\emph{USENIX
  Security 18}} (Baltimore, MD).
\newblock


\bibitem[\protect\citeauthoryear{Zhang and Papadimitratos}{Zhang and
  Papadimitratos}{2019a}]%
        {zhang2019on}
\bibfield{author}{\bibinfo{person}{K. Zhang} {and} \bibinfo{person}{P.
  Papadimitratos}.} \bibinfo{year}{2019}\natexlab{a}.
\newblock \showarticletitle{{On the Effects of Distance-decreasing Attacks on
  Cryptographically Protected GNSS Signals}}. In
  \bibinfo{booktitle}{\emph{Proceedings of ION ITM}} (Reston, VA).
\newblock


\bibitem[\protect\citeauthoryear{Zhang and Papadimitratos}{Zhang and
  Papadimitratos}{2019b}]%
        {zhang2019safe}
\bibfield{author}{\bibinfo{person}{K. Zhang} {and} \bibinfo{person}{P.
  Papadimitratos}.} \bibinfo{year}{2019}\natexlab{b}.
\newblock \showarticletitle{{Safeguarding NMA Enhanced Galileo OS Signals from
  Distance-Decreasing Attacks}}. In \bibinfo{booktitle}{\emph{Proceedings of
  ION GNSS+ 2019}} (Miami, FL). \bibinfo{pages}{4041--4052}.
\newblock


\bibitem[\protect\citeauthoryear{Zhang, Spanghero, and Papadimitratos}{Zhang
  et~al\mbox{.}}{2020}]%
        {kzmsppPLANS2020}
\bibfield{author}{\bibinfo{person}{K. Zhang}, \bibinfo{person}{M. Spanghero},
  {and} \bibinfo{person}{P. Papadimitratos}.} \bibinfo{year}{2020}\natexlab{}.
\newblock \showarticletitle{{Protecting GNSS-based Services using Time Offset
  Validation}}. In \bibinfo{booktitle}{\emph{2020 IEEE/ION PLANS}} (Portland,
  OR).
\newblock


\end{thebibliography}

\end{document}